\documentclass[aps,prd,preprintnumbers,superscriptaddress,nofootinbib,amsmath,amssymb,showpacs]{revtex4}
\usepackage{graphicx}
\usepackage{dcolumn}
\usepackage{bm}
\usepackage{amsmath}
\usepackage{color}

\input{colordvi.tex}

\newcommand{\CO}{{\cal O}}

\newcommand{\beq}{\begin{equation}}
\newcommand{\eeq}{\end{equation}}
\newcommand{\beqa}{\begin{eqnarray}}
\newcommand{\eeqa}{\end{eqnarray}}

\begin{document}

\title{On the strong coupling scale in Higgs G-inflation}

\author{Kohei Kamada}
\email[Email: ]{kohei.kamada"at"epfl.ch}
\affiliation{ Institut de Th\'eorie des Ph\'enom\`enes Physiques,
\'Ecole Polytechnique F\'ed\'erale de Lausanne, 1015 Lausanne, Switzerland}


\begin{abstract}

Higgs G-inflation is an inflation model that 
takes advantage of a Galileon-like derivative coupling. 
It is a non-renormalizable operator and is strongly coupled at
high energy scales.  Perturbative analysis does not have a predictive power any longer there. 
In general, when the Lagrangian is expanded around the vacuum, 
the strong coupling scale is identified as the mass scale that appears in non-renormalizable
operators. 
In inflationary models, however, the identification of the strong coupling scale is subtle, 
since the structures of the kinetic term as well as the interaction itself can be modified by the 
background inflationary dynamics. Therefore, the strong coupling scale depends on the background. 
In this letter, we evaluate the strong coupling scale of the fluctuations around the 
background in the Higgs G-inflation including the Nambu-Goldstone modes associated with the symmetry breaking.  
We find that the system is sufficiently weakly coupled when the scales which 
we now observe exit the horizon
during inflation and the observational predictions with the semiclassical 
treatment are valid. 
However, we also find that the inflaton field value at which the strong coupling scale and 
the Hubble scale meet is less than the Planck scale. 
Therefore, we cannot describe the model from the Planck scale, or the chaotic initial condition. 

\end{abstract}
\maketitle

\section{Introduction}

The identification of inflaton is one of the remaining key pieces in the inflationary cosmology.
Since the Higgs field is its unique candidate in the standard model (SM), 
investigating the possibilities of inflation driven by the SM Higgs field 
is one of the important issues for both high energy physics and cosmology.  
Among many proposals of the Higgs inflation models \cite{Bezrukov:2007ep,Germani:2010gm,Kamada:2012se}, 
Higgs G-inflation \cite{Kamada:2010qe}, where the kinetic term of the Higgs field is dominated by a
Galileon-like term \cite{G1,Horndeski,Deffayet:2009wt,Kobayashi:2010cm} during inflation, has a distinct characteristic. 
It can generate a large gravitational wave background,\footnote{It is also possible to generate a large gravitational wave 
background in the Higgs inflation with non-minimal coupling to gravity \cite{Bezrukov:2014bra}. } violating the (standard) Lyth bound \cite{Lyth:1996im,Kunimitsu}. 
Moreover, it breaks the standard consistency 
relation between the tensor-to-scalar ratio and the spectral tilt of the power spectrum 
of the tensor modes, $n_T= -r/8$, without introducing any non-trivial direct couplings between 
inflaton and gravity. 

The Galileon-like term that we introduce here\footnote{
Note that this term is {\it not} Galilean symmetric. 
We call it a ``Galileon-like'' term because it belongs to and is motivated by ${\cal L}_3$ 
in the generalized Galileon \cite{Deffayet:2009wt,Kobayashi:2010cm} (or the Horndeski theory \cite{Horndeski}).
}  $\phi \Box \phi (\partial \phi)^2/2M^4$ contains 
a mass parameter $M\sim 10^{13}$ GeV \cite{Kamada:2010qe}, 
which is the strong coupling scale or the ultraviolet (UV) cutoff of the theory around the vacuum. 
Inflation takes place 
at $\phi \gtrsim \sqrt{M M_{\rm pl}}$ with the Hubble parameter $H\gtrsim M$. 
Here $M_{\rm pl}(>M)$ is the reduced Planck mass and  
$\phi$ is the inflaton or the physical Higgs field in the unitary gauge. 
As is addressed in the case of the Higgs inflation with non-minimal coupling to gravity 
and its resemblances \cite{Burgess:2009ea,Bezrukov:2010jz}, 
one may wonder if the model is self-consistent and the solution 
is reliable at such a high scale. 
Since the Hubble parameter during inflation is larger than $M$, the unitarity 
of the scattering amplitude 
seems to be violated during inflation. 
However, in the inflationary background, this estimation is not correct. 
Since the structures of the kinetic term and the interaction of the fluctuation 
around the background solution differ from the ones around the vacuum, 
the strong coupling scale depends on the background dynamics 
as well as the model parameter $M$, as also discussed in, {\it e.g.}, 
Ref.~\cite{Nicolis:2004qq}, in particular in other 
Higgs inflation models  \cite{Bezrukov:2007ep,Germani:2010gm,Bezrukov:2012hx}. 
Fluctuations typically carry energies with the order of the 
Hubble scale in the quasi-de Sitter background, 
and hence the quantum fluctuations are under control 
if the Hubble parameter is smaller than the 
strong coupling scale. 

In this letter, we investigate the strong coupling scale in the Higgs G-inflation.
Here we identify the strong coupling scale as the scale where the 
tree-level unitarity is violated following the discussion in Ref.~\cite{Bezrukov:2010jz}.  
We show that in this criterion the system is sufficiently weakly coupled 
when the present horizon scales 
exited the horizon during inflation. 
Thus, the cosmological predictions evaluated in the previous studies \cite{Kamada:2012se,Kamada:2010qe} 
are valid. 
However, we also find that the (tree-level) unitarity of the scattering amplitude is violated 
at a scale less than the Planck scale, 
which suggests that the Higgs G-inflation cannot take place at the Planck scale, 
or at least we cannot describe how the Universe evolves when it starts from the Planck scale. 
Although the structure of the model is just a modification of chaotic inflation, 
the chaotic initial condition \cite{Linde:1983gd}, where Universe starts from the Planck scale, 
$K\sim V\sim M_{\rm pl}^4$, with $K$ and $V$ being the kinetic and potential energy, 
respectively, is problematic. 
Since the Higgs field value itself during inflation is larger than the strong coupling scale, 
it would be impossible to connect the results of the low-energy collider experiments 
to the precise values of running couplings at the inflationary scale 
without the knowledge of the UV physics
behind the model,
as is the case of the Higgs inflation with non-minimal coupling to gravity 
\cite{Bezrukov:2010jz,Burgess:2014lza,Bezrukov:2014ipa}.

\section{Higgs G-inflation}

First we summarize the Higgs G-inflation \cite{Kamada:2010qe}. 
It is one of the Higgs inflation models where inflation 
is driven by the potential energy of the SM Higgs field ${\cal H}$ with a Galileon-like 
derivative coupling, $({\cal H}^\dagger D_\mu D^\mu {\cal H} +{\rm h.c.})|D_\mu {\cal H}|^2/M^4$. 
The Lagrangian is given by
\begin{equation}
{\cal L}= \sqrt{-g} \left[\frac{M_{\rm pl}^2}{2}R-|D_\mu {\cal H}|^2 -\left(\frac{{\cal H}^\dagger}{M^4}D_\mu D^\mu {\cal H}+{\rm h.c.}\right) |D_\mu {\cal H}|^2 -\lambda |{\cal H}|^4\right], 
\end{equation}
where $R$ is the Ricci scalar, 
$D_\mu$ is the covariant derivative, and $M$ and $\lambda$ are model parameters 
whose mass dimensions are 1 and 0, respectively. 
Here we omit the Higgs mass term since it is irrelevant to the inflationary dynamics. 
We adopt the Friedman-Robertson-Walker metric, $ds^2=-dt^2+a(t)^2 \delta_{ij} dx^i dx^j$, 
with the Hubble parameter $H={\dot a}/a$. 
Let us first investigate the physical Higgs in the unitary gauge, ${\cal H}=(0,\phi)/\sqrt{2}$. 
The Lagrangian for $\phi$ is given by
\begin{equation}
{\cal L}=\sqrt{-g}\left[ \frac{M_{\rm pl}^2}{2}R -\frac{1}{2}(D_\mu \phi)^2-\frac{\phi D_\mu D^\mu \phi}{2M^4}(D_\mu \phi)^2-\frac{\lambda}{4}\phi^4\right]. 
\end{equation}
This system allows a potential-driven slow-roll inflation when the slow-roll conditions, 
$|\epsilon|, |\eta|, |\alpha| \ll 1$,  are satisfied. Here the slow-roll parameters are given by
\begin{equation}
\epsilon \equiv -\frac{\dot H}{H^2}, \ \ \ \eta \equiv -\frac{\ddot \phi}{H{\dot \phi}},\ \ \  \alpha \equiv \frac{\dot \phi}{H \phi}. 
\end{equation}
Note that we here introduced a new slow-roll parameter $\alpha$ to take into account 
the effect of the Galileon-like derivative coupling \cite{Kamada:2010qe}. 
The slow-roll equations are found to be 
\begin{align}
&3H^2 M_{\rm pl}^2=\frac{\lambda}{4}\phi^4, \label{SL1}\\
&3H{\dot \phi}\left(1-\frac{3H {\dot \phi}\phi}{M^4}\right)+\lambda \phi^3=0. \label{SL2}
\end{align}
Here we assumed that the effect of running of the model parameters is negligible 
and they are taken as constants. 
In particular, we focus on the case where $\lambda=\CO(10^{-2})>0$ at the inflationary scale.\footnote{
Though the recent experimental results \cite{Aad:2012tfa} suggest the metastability of the Higgs potential \cite{Bezrukov:2012sa}, 
the parameter space where the Higgs quartic coupling is positive around the scale 
of the grand unified theory (GUT) is not excluded.}
For $\phi>\sqrt{2} \lambda^{-1/4} M$, the second term in the parenthesis in Eq.~\eqref{SL2}
dominates the first term, and we have the inflationary solution
\begin{equation}
{\dot \phi}=-\frac{2 M^2 M_{\rm pl}}{\sqrt{3}\phi}.
\end{equation}
Inflation ends when the slow-roll condition $|\epsilon| \ll 1$ breaks, 
\begin{equation}
\phi=\phi_{\rm end}\equiv 2^{3/4} \lambda^{-1/8} \sqrt{M M_{\rm pl}}. 
\end{equation}
The Higgs field value at the number of $e$-folds ${\cal N}$ before the end of 
inflation is evaluated as
\begin{equation}
\phi_{\cal N}=(16 {\cal N}+8)^{1/4} \lambda^{-1/8} \sqrt{M M_{\rm pl}}. 
\end{equation}
The power spectrum of the primordial scalar perturbation $A_s$ is calculated as
\begin{equation}
A_s=\frac{(2{\cal N}+1)^2}{8\pi^2}\left(\frac{3}{8}\right)^{1/2}\lambda^{1/2}\left(\frac{M}{M_{\rm pl}}\right)^2, 
\end{equation}
and its spectral tilt is evaluated as 
\begin{equation}
n_s=1-\frac{4}{2 {\cal N}+1}. 
\end{equation}
For the number of $e$-folds ${\cal N}_* \simeq 60$ where 
the pivot scale $k_0=0.05 {\rm Mpc}^{-1}$ exited the horizon during inflation, the measurement of the Planck satellite,   
$A_s \simeq 2.2 \times 10^{-9}$ \cite{Planck:2013jfk}, is reproduced by $M \simeq (\lambda/0.01)^{-1/4}\times 3 \times 10^{13}$ GeV. 
Consequently, during inflation the Higgs field value is around the GUT scale. 
The tensor-to-scalar ratio is given by\footnote{
This value of the tensor-to-scalar ratio is disfavored by the recent results of Planck. 
However, here we neglect the effect of the running coupling constant, 
which may be able to reduce it slightly to reconcile with the observation. }
\begin{equation}
r=\frac{64}{3}\left(\frac{2}{3}\right)^{1/2} \epsilon \simeq \frac{17}{2 {\cal N}_*+1}\sim 0.14.  
\end{equation}
Since the model does not change the tensor sector compared to the standard scenario, 
the tensor spectral tilt is expressed by the slow-roll parameter as the standard 
expression $n_T\simeq -2\epsilon$. 
As a result, we have a non-standard consistency relation, 
\begin{equation}
r\simeq -\frac{32 \sqrt6}{9} n_T. 
\end{equation}

\section{Strong coupling in the inflaton self-interaction}
One may wonder if this inflationary solution gives a consistent scenario. 
If the fluctuations around the inflationary trajectory couple too strongly, 
or the strong coupling scale (or the cutoff scale) is smaller than the Hubble scale, 
the above discussion is no longer reliable. 
In order to determine it, here we adopt the discussion in Ref.~\cite{Bezrukov:2010jz} 
(see also Ref.~\cite{Nicolis:2004qq} for the similar discussion in the DGP model),  
where we identify the cutoff scale as the scale where the tree-level unitarity  is violated 
\cite{Cornwall:1974km}. 
Here we divide the Higgs field into the slowly varying classical part and excitations and 
estimate the cutoff scale by the power counting of the operators in the action 
expanded with respect to the canonically normalized excitations ${\tilde \chi}_i$.\footnote{
The subscript $i$ is introduced to take into account the Nambu-Goldstone (NG) modes.} 
Once the operators are expanded as
\begin{equation}
\frac{{\cal O}_{(n)} (\chi_i, \partial \chi_i, \partial^2 \chi_i)}{[\Lambda_{(n)}({\bar \phi}, {\dot {\bar \phi}})]^{n-4}}, 
\end{equation}
we expect that the tree-level unitarity is violated at $E_{\rm sc}={\rm min.}\{\Lambda_{(n)}\}$. 
If it is sufficiently larger than the Hubble parameter during inflation, 
the semiclassical treatments of the model are valid and it is self-consistent. 

First we see the self-interaction of the fluctuation along the inflationary trajectory. 
Let us rewrite the inflaton field as
\begin{equation}
\phi (x) \rightarrow {\bar \phi}(t)+\chi(x),  
\end{equation}
where ${\bar \phi}(t)$ is the homogeneous background solution and $\chi(x)$
is the fluctuation around the background. 
Noting that 
\begin{align}
D_\mu D^\mu \phi &=-{\ddot {\bar \phi}}-3H{\dot {\bar \phi}}+\Box \chi,  &
(D_\mu \phi)^2 &=-{\dot {\bar \phi}}^2-2 {\dot {\bar \phi}}{\dot \chi}+(D_\mu \chi)^2, 
\end{align}
we can expand the derivative coupling as
\begin{align}
-\frac{1}{2}(D_\mu\phi)^2&-\frac{1}{2M^4}\phi \Box \phi (D_\mu \phi)^2 \notag \\
&=\frac{1}{2} \left(1-\frac{{\bar \phi}({\ddot {\bar \phi}}+3H{\dot {\bar \phi}})}{M^4}\right){\dot {\bar \phi}}^2-\frac{({\ddot {\bar \phi}}+3H{\dot {\bar \phi}}){\dot {\bar \phi}}^2}{2M^4} \chi+\left(1-\frac{{\bar \phi}({\ddot {\bar \phi}}+3H{\dot {\bar \phi}})}{M^4}\right){\dot {\bar \phi}} {\dot \chi}+\frac{{\bar \phi}{\dot {\bar \phi}}^2}{2M^4}\Box \chi \notag \\
&+\frac{{\dot {\bar \phi}}^2}{2M^4}\chi \Box \chi-\frac{1}{2}\left(1-\frac{{\bar \phi}({\ddot {\bar \phi}}+3H{\dot {\bar \phi}})}{M^4}\right)(D_\mu \chi)^2-\frac{{\dot {\bar \phi}}({\ddot {\bar \phi}}+3H{\dot {\bar \phi}})}{M^4}\chi{\dot \chi}+\frac{{\bar \phi}{\dot {\bar \phi}}}{M^4}{\dot \chi}\Box \chi \notag \\
&+\frac{{\dot {\bar \phi}}}{M^4}\chi {\dot \chi}\Box \chi+\frac{{\ddot {\bar \phi}}+3H{\dot {\bar \phi}}}{2M^4}\chi (D_\mu \chi)^2-\frac{\bar \phi}{2M^4}\Box \chi (D_\mu \chi)^2-\frac{1}{2M^4}\chi \Box \chi (D_\mu \chi)^2.
\end{align}
With taking a partial integral, the kinetic term of $\chi$ is given by
\begin{align}
S_{\rm kin}[\chi]=\int d^4 x a^3(t) &\left(\frac{{\dot {\bar \phi}}^2}{2M^4}\chi \Box \chi-\frac{1}{2}\left(1-\frac{{\bar \phi}({\ddot {\bar \phi}}+3H{\dot {\bar \phi}})}{M^4}\right)(D_\mu \chi)^2+\frac{{\bar \phi}{\dot {\bar \phi}}}{M^4}{\dot \chi}\Box \chi \right)\notag \\
=\int d^4 x a^3(t)&\left[\frac{1}{2}\left(1+\frac{2{\dot {\bar \phi}}^2-6H{\bar \phi}{\dot {\bar \phi}}}{M^4}\right){\dot \chi}^2-\frac{1}{2}\left(1-\frac{2{\bar \phi}({\ddot {\bar \phi}}+2H{\dot {\bar \phi}})}{M^4}\right)\frac{\delta^{ij}}{a^2(t)}\partial_i \chi \partial_j \chi \right] \notag \\
&+(\text{mass term})+(\text{total derivative}). 
\end{align}
The inflaton fluctuation can be canonically normalized as follows. 
Defining
\begin{equation}
G(t) \equiv 1+\frac{2{\dot {\bar \phi}}^2-6H{\bar \phi}{\dot {\bar \phi}}}{M^4}, \ \ F(t)\equiv 1-\frac{2{\bar \phi}({\ddot {\bar \phi}}+2H{\dot {\bar \phi}})}{M^4}
\end{equation}
and 
\begin{equation}
d {\tilde x}^i \equiv \sqrt{\frac{G(t)}{F(t)}}a(t)dx^i, \quad 
{\tilde \chi}\equiv\frac{F(t)^{3/4}}{G(t)^{1/4}} \chi,
\end{equation}
we see that $\tilde \chi$ is canonically normalized, 
\begin{equation}
S_{\rm kin}[\chi]=\int dt d^3{\tilde x} \frac{1}{2} \left({\dot {\tilde \chi}}^2-\delta^{ij}\frac{\partial {\tilde \chi}}{\partial {\tilde x}^i}\frac{\partial {\tilde \chi}}{\partial {\tilde x}^i}\right)
+(\text{mass term})+(\text{total derivative}). 
\end{equation}
Note that there arise terms coming from the derivatives of $F(t)$ 
and $G(t)$, but they are mass terms and total derivatives, and hence we do not 
write explicitly. 
With the above coordinate and field redefinition, 
higher derivative interactions are rewritten as
\begin{align}
S_{\rm int}=\int dt d^3 {\tilde x} &\frac{1}{G(t)^{3/4}F(t)^{3/4}}\frac{\dot {\bar \phi}}{M^4}{\tilde \chi}{\dot {\tilde \chi}}\Box {\tilde \chi}+\frac{1}{G(t)^{3/4}F(t)^{3/4}}\frac{{\ddot {\bar \phi}}+3H{\dot {\bar \phi}}}{2M^4}{\tilde \chi} (D_\mu {\tilde \chi})^2\notag \\
&-\frac{1}{G(t)^{3/4}F(t)^{3/4}}\frac{\bar \phi}{2M^4}\Box {\tilde \chi} (D_\mu {\tilde \chi})^2-\frac{1}{G(t)^{1/2}F(t)^{3/2}}\frac{1}{2M^4}\chi \Box \chi (D_\mu \chi)^2. \label{21}
\end{align}
Here we again omit the terms coming from the derivatives of $F(t)$ 
and $G(t)$. But these terms are slow-roll suppressed, and hence they are less 
dangerous than the terms in Eq.~\eqref{21} in the inflationary background. 
We also do not write the modification of spatial derivative explicitly, 
which changes the structure of 
$\Box {\tilde \chi}$ and $(D_\mu {\tilde \chi})^2$ slightly since $F(t) \sim G(t)$, and hence 
the phase velocity is of the order of the unity if there are not any non-trivial cancelations, 
(which is true for the slow-roll inflationary solution as we will see below). 
Consequently, we identify the strong coupling scale for the $\tilde \chi$ field as 
\begin{equation}
E>E_{\rm sc}\equiv {\rm min}. \left\{\frac{G(t)^{3/8}F(t)^{3/8}M^2}{{\dot {\bar \phi}}^{1/2}}, \frac{2G(t)^{3/4}F(t)^{3/4}M^4}{{\ddot {\bar \phi}}+3H{\dot {\bar \phi}}}, \frac{2^{1/3}G(t)^{1/4}F(t)^{1/4}M^{4/3}}{{\bar \phi}^{1/3}}, 2^{1/4} G(t)^{1/8}F(t)^{3/8}M\right\}. 
\end{equation}

During  inflation, we have the slow-roll trajectory, 
\begin{equation}
H\simeq \frac{\lambda^{1/2}{\bar \phi}^2}{2\sqrt{3}M_{\rm pl}}, \quad {\dot {\bar \phi}}=-\frac{2 M^2 M_{\rm pl}}{\sqrt{3} {\bar \phi}}, \quad {\ddot {\bar \phi}}=-\frac{4 M^4 M_{\rm pl}^2}{3 {\bar \phi}^3}, \quad |H{\dot {\bar \phi}}| \gg |{\ddot {\bar \phi}}|, 
\end{equation}
and hence
\begin{equation}
G(t) \simeq -\frac{6H{\bar \phi}{\dot {\bar \phi}}}{M^4} \simeq \frac{2 \lambda^{1/2} {\bar \phi}^2}{M^2},\quad F(t) \simeq -\frac{4H{\bar \phi}{\dot {\bar \phi}}}{M^4} \simeq \frac{4 \lambda^{1/2} {\bar \phi}^2}{3M^2}. 
\end{equation}
As a result, we find that the strong coupling scale is given by
\begin{equation}
E_{\rm sc} ({\bar \phi})=\frac{2^{1/3}G(t)^{1/4}F(t)^{1/4} M^{4/3}}{{\bar \phi}^{1/3}}\simeq \lambda^{1/4} {\bar \phi}^{2/3} M^{1/3}
\end{equation}
for ${\bar \phi}>\phi_{\rm end}\sim \sqrt{M M_{\rm pl}}$. 
Therefore, for 
\begin{equation}
H({\bar \phi})\gg E_{\rm sc}({\bar \phi}) \Leftrightarrow {\bar \phi} \gg \lambda^{-3/16}M^{1/4} M_{\rm pl}^{3/4} \simeq 0.1 \left(\frac{\lambda}{0.01}\right)^{-3/16}\left(\frac{M}{10^{13}{\rm GeV}}\right)^{1/4} M_{\rm pl} \equiv {\bar \phi}_{\rm sc},  
\end{equation}
the system is strongly coupled compared to the inflationary scale. 
Since $\phi_{{\cal N}_*}\ll {\bar \phi}_{\rm sc}$ for ${\cal N}_*\simeq 60$, 
the $\chi$ field is weakly self-interacted when the scale we now see in the CMB
exited the horizon during inflation. 
This constraint 
also suggests that the Higgs G-inflation cannot start from the Higgs field value larger 
than the Planck 
scale, unlike the usual chaotic inflation. 
Note that for
\begin{equation}
H^2({\bar \phi}) \ll \frac{F({\bar \phi})^{3/4}}{G({\bar \phi})^{1/4}}{\dot {\bar \phi}} \Leftrightarrow {\bar \phi}\ll \lambda^{-3/16} M^{1/4} M_{\rm pl}^{3/4} \simeq{\bar \phi}_{\rm sc}, 
\end{equation}
quantum fluctuations are small compared with the classical evolution of the inflaton. 
Therefore the Higgs G-inflation cannot have the stage of eternal inflation. 

Note that here we neglect the mixing between the inflaton fluctuation and
scalar perturbation of the metric tensor. However, they are decoupled 
in the $M_{\rm pl} \rightarrow \infty$ limit with $\lambda \phi^4/M_{\rm pl}^2 = \text{const}$. 
We can see that the quadratic action of the scalar metric perturbation in the unitary gauge 
\cite{Kamada:2010qe} differs from the action of 
inflaton fluctuations only by Planck suppressed terms, 
which are subdominant during inflation.
Therefore, the mixing between the inflaton fluctuation and 
the scalar perturbation of the metric tensor is negligible for our purposes.

\section{Strong coupling in the Nambu-Goldstone sector}
Since the transverse modes of gauge bosons and gravitons do not change 
their kinetic term by the Galileon-like Higgs derivative coupling, 
the strong coupling scale in and between these sectors is equal to or larger 
than the one in the Higgs self-interaction. 
However, the kinetic terms of the longitudinal mode of the gauge bosons or the NG modes 
are modified by the Galileon-like derivative coupling. 
Therefore we cannot tell if the system is weakly coupled during inflation 
unless we also check the interaction of NG modes. 
Here we investigate the interaction of the NG modes along the inflationary trajectory. 

Let us expand the Higgs field along the inflationary trajectory as
\begin{equation}
{\cal H}=\frac{1}{\sqrt{2}}\left(
\begin{array}{cc}
\theta_2+i\theta_3 \\
{\bar \phi}(t)+\chi(x)+i \theta_1(x)
\end{array}\right). 
\end{equation}
Here $\chi$ is the inflaton fluctuation and $\theta_i$ are the NG modes. 
By taking a partial integral, the action for the inflaton fluctuations $\chi$ 
and the NG modes $\theta_i$ is written by
\begin{align}
S_{\rm kin}=\int d^4 x a^3(t)&\left[\frac{1}{2}\left(1+\frac{2{\dot {\bar \phi}}^2-6H{\bar \phi}{\dot {\bar \phi}}}{M^4}\right){\dot \chi}^2-\frac{1}{2}\left(1-\frac{2{\bar \phi}({\ddot {\bar \phi}}+2H{\dot {\bar \phi}})}{M^4}\right)\frac{\delta^{ij}}{a^2(t)}\partial_i \chi \partial_j \chi\right. \notag \\
&-\frac{1}{2}\left(1-\frac{{\bar \phi}{\ddot {\bar \phi}-{\dot {\bar \phi}}^2+3H{\bar \phi}{\dot {\bar \phi}}}}{M^4}\right)\left(\sum_i(\partial_\mu \theta_i)^2\right)\notag \\
&+\frac{1}{M^4}({\ddot {\bar \phi}} \chi-{\bar \phi} \partial_\mu \partial^\mu \chi)\left((\partial_\mu \chi)^2+\sum_i(\partial_\mu \theta_i)^2\right)+\frac{{\dot {\bar \phi}}}{M^4}{\dot \chi}\left(\chi \partial_\mu \partial^\mu \chi+\sum_i \theta_i \partial_\mu \partial^\mu \theta_i\right) \notag \\
&\left.-\frac{1}{2M^4}\left((\partial_\mu \chi)^2+\sum_i(\partial_\mu \theta_i)^2\right)\left(\chi \partial_\mu \partial^\mu \chi+\sum_i \theta_i \partial_\mu \partial^\mu \theta_i\right)  \right]+\text{(potential terms)} + \text{(total derivative)}.
\end{align}
During inflationary stage, we have $3H{\bar \phi} {\dot {\bar \phi}} \gg, {\dot {\bar \phi}}^2, {\bar \phi}{\ddot {\bar \phi}}, M^4$, and hence the action is approximated as 
\begin{align}
S_{\rm kin}\simeq \int d^4 x a^3(t)&\left[-\frac{1}{2}\frac{6H{\bar \phi}{\dot {\bar \phi}}}{M^4}\left({\dot \chi}^2-\frac{2}{3}\frac{\delta^{ij}}{a^2(t)}\partial_i \chi \partial_j \chi\right)\right. +\frac{1}{2}\frac{3H{\bar \phi}{\dot {\bar \phi}}}{M^4}\left(\sum_i\left({\dot \theta}_i^2-\frac{\delta^{jk}}{a^2(t)}\partial_j \theta_i \partial_k \theta_i\right)\right)\notag \\
&+\frac{1}{M^4}({\ddot {\bar \phi}} \chi-{\bar \phi} \partial_\mu \partial^\mu \chi)\left((\partial_\mu \chi)^2+\sum_i(\partial_\mu \theta_i)^2\right)+\frac{{\dot {\bar \phi}}}{M^4}{\dot \chi}\left(\chi \partial_\mu \partial^\mu \chi+\sum_i \theta_i \partial_\mu \partial^\mu \theta_i\right) \notag \\
&\left.-\frac{1}{2M^4}\left((\partial_\mu \chi)^2+\sum_i(\partial_\mu \theta_i)^2\right)\left(\chi \partial_\mu \partial^\mu \chi+\sum_i \theta_i \partial_\mu \partial^\mu \theta_i\right) \right] +\text{(potential terms)}+ \text{(total derivative)}.
\end{align}
We can see that the NG modes have almost the same kinetic term structures 
to that of the inflaton fluctuation. 
Consequently, NG modes are canonically normalized with subluminal sound 
speeds  
by almost the same field redefinition to the inflaton fluctuations. 
As a result, the strong coupling scale in the NG boson sector 
is also the same to the $\chi$ self-interaction, $i.e.$, 
\begin{equation}
E_{\rm sc} ({\bar \phi})\simeq \lambda^{1/4} {\bar \phi}^{2/3} M^{1/3}. 
\end{equation}
We here conclude that the field dependent cutoff scale of the Higgs G-inflation model 
estimated by the power counting of all the fluctuation operators is 
sufficiently larger than the energy scale carried by the fluctuations. 
Therefore, the requirement for the validity of the semiclassical treatment 
performed in the previous studies \cite{Kamada:2012se,Kamada:2010qe} is fulfilled. 
Note that the derivative interactions between the Higgs and the NG modes induce 
ghost-like degrees of freedom \cite{Koehn:2013hk}, but it is harmless since the scale at which 
they appear is also the 
strong coupling scale, which is larger than the Hubble parameter during inflation.  

\section{Summary and Discussion}

In this letter, we evaluate the strong coupling scale in the Higgs G-inflation 
identifying it as the scale where the tree-level unitarity is violated. 
We find that the strong coupling scale of the fluctuations around the inflationary trajectory
depends on the background dynamics and 
is larger than the Hubble scale when the present horizon scale exited the horizon. 
Since the inflationary background modifies the structure of the kinetic term 
of the fluctuations, the mass parameter in the original Lagrangian 
is not directly related to the strong coupling scale.  
As a result, the semiclassical calculation performed in the previous studies 
\cite{Kamada:2012se,Kamada:2010qe} are valid and the model is self-consistent. 
Note that the strong coupling scale meets the Hubble scale at 
${\bar \phi} \simeq \lambda^{-3/16} M^{1/4} M_{\rm pl}^{3/4} (< M_{\rm pl})$. 
This suggests that we cannot describe the onset of the Higgs G-inflation from 
the chaotic initial condition, where both potential and kinetic energies are 
of the order of the Planck scale.
For the related study on the generalized G-inflation \cite{Kobayashi:2010cm}, 
see Ref.~\cite{Kunimitsu}. 

Note that the condition that we adopt in this letter is the necessary condition, but not the 
sufficient condition \cite{Bezrukov:2010jz}. 
Since the Galileon-like derivative coupling is a non-renormalizable operator, 
quantum corrections generate an infinite number of higher order interactions. 
To remove the divergences, we need to add an infinite number of counter terms. 
In order to tell the sufficient condition of the validity of the model, 
we need to examine how these loop corrections can be suppressed, 
but it is a matter of ``naturalness''. 
In the case of the Higgs inflation with non-minimal coupling to gravity, the theory has an 
asymptotic scale invariance, which guarantees the absences of higher order terms, 
and hence it is ``natural'' \cite{Bezrukov:2010jz}. 
The naturalness in $P(X)$ and Galileon theories 
that have additional symmetries like Galilean symmetry or shift symmetry 
is studied in Ref.~\cite{deRham:2014wfa}. 
However, the higher derivative term we introduce in the Higgs G-inflation 
does not have known (asymptotic) symmetry behind it. 
Note that the ``Galileon-like'' term as well as the potential term is not Galilean symmetric. 
Nevertheless, the fact that the strong coupling scale is large enough for the Higgs G-inflation
may suggest that the existence of a hidden asymptotic symmetry, which makes the model ``natural''.  But further investigations 
are needed which is left for the future study. 

Apart from the naturalness, what we find here has an important insight in the connection 
between the inflationary parameters and low-energy physics. 
Since the Higgs field value during inflation is larger than the
strong coupling scale, there can appear 
higher order interactions to modify the running of the couplings or 
at least threshold effect between the electroweak scale and the inflationary scale, 
as are discussed in Refs.~\cite{Bezrukov:2010jz,Burgess:2014lza,Bezrukov:2014ipa}. 
Therefore, their connection is sensitive to the detail of the UV completion
of the theory. 
In particular, even if the low-energy experiments will suggest the meta-stability of the 
Higgs potential \cite{Bezrukov:2012sa,Espinosa:2007qp}, the Higgs G-inflation may be still possible 
depending on the UV completion as is the case discussed recently in Ref.~\cite{Bezrukov:2014ipa}. 
Note that there are discussions on the UV completion of the (generalized) Galileons. 
Since the (generalized) Galileon theories exhibit low-energy superluminalities of fluctuations in specific 
backgrounds, 
it may suggest the absence of any local Lorentz invariant UV completions \cite{Adams:2006sv}. 
However, it is still a subtle issue \cite{Creminelli:2013fxa,Hollowood:2007kt,Dubovsky:2007ac}, 
and hence careful studies on the UV completion are needed for the detailed investigation 
of the Higgs G-inflation. 

\section*{Acknowledgments}
The author is grateful to S.~Sibiryakov for helpful discussions and comments. 
The author also thanks T.~Kunimitsu and J.~Yokoyama for useful comments. 
This work has been supported in part by the JSPS Postdoctoral Fellowships for Research Abroad.

\end{document}